\documentclass[aps,prb,twocolumn]{revtex4} 

\usepackage{graphicx}
\usepackage{dcolumn}
\usepackage{bm}

\newcommand{\comment}[1]{}

\begin{document}

\title{Quantum Statistical Mechanics. I.
Decoherence, Wave Function Collapse, and the von Neumann Density Matrix}


\author{Phil Attard}

\date{31 Oct., 2013,\ phil.attard1@gmail.com}

\begin{abstract}
The probability operator is derived from first principles
for an equilibrium  quantum system.
It is also shown that
the  superposition states collapse into a mixture of states
giving the conventional von Neumann trace form for the quantum average.
The mechanism for the collapse is found to be quite general:
it results from the conservation law for a conserved, exchangeable variable
(such as energy) and the entanglement of the total system wave function
that necessarily follows.
The relevance of the present results
to the einselection mechanism for decoherence,
to the quantum measurement problem,
and to the classical nature of the macroscopic world
are discussed.

\end{abstract}


\maketitle

                \section{Introduction}

Perhaps the most puzzling aspect of quantum mechanics
is that a quantum system may exist simultaneously in a superposition
of states, whereas in the classical world a system can only ever be in
one state at a time.
The precise  process whereby a quantum system
with its allowed superposition of states
passes to the observed classical limit of pure states
is the subject of ongoing debate;
Schr\"odinger's cat is periodically resuscitated
for \emph{post mortem} examination.

In quantum statistical mechanics,
the superposition problem disappears
in the conventional formulation
in terms of the von Neumann density matrix.
\cite{Neumann27,Messiah61,Merzbacher70,Bogulbov82}
This expresses the statistical aspects of the problem
as an ensemble or mixture of systems, each in a pure quantum state,
which is akin to the classical experience of one  state at a time.

Two fundamental questions naturally arise:
how does one get from the underlying quantum superposition wave function
to a mixture of  quantum states,
and how does one derive the statistical probability distribution
from quantum mechanical first principles?
Both questions are addressed in this paper.

The appearance of a mixture of states from a superposition of states
is referred to here as the collapse of the system or wave function.
This is of course closely related to the traditional
use of the word `collapse' in reference
to the transition of a quantum system from a superposition of states
to a single eigenstate following the application of an operator.
The so-called measurement problem,
the emergence of the mixed state von Neumann density matrix,
and the classical nature of the macroscopic world
are all closely related phenomena.

One promising approach to the problems of wave function collapse
and the classical behavior of macroscopic objects
explores the decoherence that arises from the interactions
of a sub-system with its environment.
\cite{Zurek82,Joos85,Zurek91,Zurek94,Zeh01,Zurek03,Zurek04,Schlosshauer05}
Essentially the approach says that
the environment or reservoir
coupled to the sub-system
causes the wave function to collapse
into a mixture of pure quantum states
(environmental selection, or einselection).
That is, the superposed components
of the original sub-system wave function
are rendered decoherent by the interaction with the reservoir
or measuring device.
A brief summary of the conventional einselection approach to decoherence
and a discussion of the similarities and differences
with the present approach is given in Appendix  \ref{Sec:ein}.

The present paper derives a formula for the statistical average
of a canonical equilibrium system
(i.e.\ a sub-system that can exchange energy with a reservoir).
It is shown that due to the conservation law the
wave function of the total system necessarily
has an entangled form.
It is this that causes the system to collapse from a superposition of states
to a mixture of pure states.
Both the Maxwell-Boltzmann probability operator
and the von Neumann density matrix emerge from the first
principles derivation.

                \section{Derivation}

It is axiomatic that
\begin{equation}
\begin{array}{c}
\mbox{all microstates of an isolated system} \\
\mbox{have equal weight.}
\end{array}
\end{equation}
See Appendix \ref{Sec:weight} for a justification.
This axiom is implied by the standard
interpretation of quantum mechanics:
representing the system in the basis corresponding to some operator,
$| \psi \rangle =
\sum_{kg} \psi_{kg}^O | \zeta_{kg}^O \rangle$,
with $\hat O | \zeta_{kg}^O \rangle
= O_k | \zeta_{kg}^O \rangle$,
the product $\psi_{kg}^O \psi_{kg}^{O*}$
is interpreted as being proportional to the probability
of the system being in the microstate $\zeta_{kg}^O$.
\cite{Messiah61,Merzbacher70}
With this, the expectation value of the operator is simply
$O(\psi) = \sum_{kg} \psi_{kg}^{O*} \psi_{kg}^O O_k
/\sum_{kg} \psi_{kg}^{O*} \psi_{kg}^O $.
If the microstates did not have equal weight,
then this expression for the expectation value would need to include
an additional factor proportional to the weight of each microstate.

Consider an isolated sub-system
with Hamiltonian operator $\hat {\cal H}$
and orthonormal energy eigenfunctions such that
$ \hat {\cal H} | \zeta^\mathrm{E}_{kg} \rangle
= E_k | \zeta^\mathrm{E}_{kg} \rangle $.
Here $k$ is the principle energy quantum number
and labels the energy macrostates,
and $g=1, 2, \ldots , N_k$ labels the degenerate energy states;
$\zeta^\mathrm{E}_{kg}$ labels the microstates of the system
in the energy representation.

From the equal weight condition,
the weight of the energy microstates can be taken to be
$w_{kg}^\mathrm{E} = 1$ and the entropy of the energy microstates is
$S_{kg}^\mathrm{E} = k_\mathrm{B} \ln w_{kg}^\mathrm{E} = 0$,
where $k_\mathrm{B}$ is Boltzmann's constant.
Note that the same value is used
irrespective of the energy of the isolated system
and whether or not this is equal to $E_k$.
According to the set theoretic formulation of probability
for statistical mechanics,\cite{TDSM,NETDSM}
the weight of an energy macrostate is therefore
$w_{k}^\mathrm{E} = \sum_g w_{kg}^\mathrm{E} = N_k$
and the entropy of an energy macrostate
is just the logarithm of the number of microstates that it contains,
$S_{k}^\mathrm{E} = k_\mathrm{B} \ln N_k $.
This is Boltzmann's original formula.
Finally, soon will be invoked the thermodynamic definition of temperature,
namely\cite{TDSM,NETDSM}
\begin{equation}
\frac{1}{T} = \frac{\partial S(E)}{\partial E} .
\end{equation}

Now allow the sub-system s to exchange energy with a reservoir r
such that the total system is isolated and therefore
$E_\mathrm{tot} = E_\mathrm{s} + E_\mathrm{r}$ is constant.
The  Hilbert space is $\mathrm{H}_\mathrm{tot}
= \mathrm{H}_\mathrm{s} \otimes \mathrm{H}_\mathrm{r}$.
If it is possible to factor the wave state of the total system as
\begin{equation}
\left| \psi_\mathrm{tot} \right>
=
\left| \psi_\mathrm{s} \right>
\left| \psi_\mathrm{r} \right> ,
\end{equation}
then the wave state is said to be a separable state,
or a product state.
Here and below the direct product symbol on the right hand side
is suppressed.
Conversely,
with $\{ | \zeta^\mathrm{s}_i \rangle \}$
a basis for the sub-system
and $\{ | \zeta^\mathrm{r}_j \rangle \}$ a basis for the reservoir,
the most general wave state of the total system is
\begin{equation}
| \psi_\mathrm{tot} \rangle
=
\sum_{i,j} c_{ij}
| \zeta^\mathrm{s}_i \rangle
| \zeta^\mathrm{r}_j \rangle .
\end{equation}
If the coefficient matrix is dyadic,
$c_{ij} = c_i^\mathrm{s} c_j^\mathrm{r}$,
then the wave state is separable with
$| \psi_\mathrm{s} \rangle
= \sum_i\!\!\!^\mathrm{(s)} c_i^\mathrm{s} | \zeta^\mathrm{s}_i \rangle$,
and
$| \psi_\mathrm{r} \rangle
= \sum_j\!\!\!^\mathrm{(r)} c_j^\mathrm{r} | \zeta^\mathrm{r}_j \rangle$.
Alternatively, if the coefficients can't be written
in dyadic form, then the wave state is inseparable.
This is called an entangled state.\cite{Messiah61,Merzbacher70}

Whether or not the total system can be written as a product state or
is necessarily an entangled state
is the main difference between the present analysis
and  the conventional decoherence einselection
approach.
\cite{Zurek82,Joos85,Zurek91,Zurek94,Zeh01,Zurek03,Zurek04,Schlosshauer05}
(See  Appendix \ref{Sec:ein}.)

In general conservation laws give rise to entangled states.
This may be seen explicitly for the present case
of energy exchange,
with $E_\mathrm{tot} = E_\mathrm{s} + E_\mathrm{r}$ fixed.
Using the energy eigenfunctions described above
as a basis for the sub-system and the reservoir,
the most general expansion of the total wave function is
\begin{equation}
| \psi_\mathrm{tot} \rangle
=
\sum_{ig,jh} c_{ig,jh}
| \zeta^\mathrm{s,E}_{ig} \rangle
| \zeta^\mathrm{r,E}_{jh} \rangle .
\end{equation}
In view of energy conservation one has,
\begin{equation} \label{Eq:c=0}
c_{ig,jh} = 0
\mbox{ if }
E_j^\mathrm{r} + E_i^\mathrm{s} \ne E_\mathrm{tot} .
\end{equation}

It is not possible to satisfy this if $c_{ig,jh}$ is dyadic.
The proof is by contradiction.
Suppose that for $i_1$, $ E_{i_1}^\mathrm{s} < E_\mathrm{tot} $
and hence there exists a corresponding $j_1$ such that
$E_{j_1}^\mathrm{r} = E_\mathrm{tot} - E_{i_1}^\mathrm{s} $.
Assuming  that the coefficient matrix is dyadic,
$c_{ig,jh} = c_{ig}^\mathrm{s} c_{jh}^\mathrm{r}$,
then for all $j \ne {j_1}$ one must have $c_{jh}^\mathrm{r} = 0$.
But this would mean that there is only one  value
of the reservoir energy,
and hence that $ E_{i_1}$ is the only possible value of the sub-system energy.
Since this contradicts the fundamental definition of the statistical system,
namely that the sub-system can exchange energy with the reservoir,
one concludes that the coefficient matrix cannot be dyadic
and that the wave state of the total system must be entangled.
\cite{FN2}

Using the entangled form of the total wave function,
and the vanishing of the coefficients for terms
that violate  energy conservation,
the expectation value of an arbitrary operator on the sub-system is
\begin{eqnarray}  \label{Eq:O(psi-tot)}
O(\psi_\mathrm{tot})
& = &
\frac{1}{Z'}
\sum_{ig,i'g'}\!\!\!^\mathrm{(s)} \sum_{jh,j'h'}\!\!\!^\mathrm{(r)}
\; c_{i'g',j'h'}^* c_{ig,jh}
\nonumber \\ && \mbox{ } \times
\langle \zeta^\mathrm{s,E}_{i'g'} |
\hat O
| \zeta^\mathrm{s,E}_{ig} \rangle
\langle \zeta^\mathrm{r,E}_{j'h'} | \zeta^\mathrm{r,E}_{jh} \rangle
\nonumber \\ & = &
\frac{1}{Z'} \sum_{ig,i'g'}\!\!\!^\mathrm{(s)}  \sum_{jh}\!\!^\mathrm{(r)}
\; c_{i'g',jh}^* c_{ig,jh}
O^\mathrm{s,E}_{i'g',ig}
\nonumber \\ & = &
\frac{1}{Z'}  \sum_{i,g,g'}\!\!\!^\mathrm{(s)}  \sum_{h}\!\!^\mathrm{(r)}
\; c_{ig',j_ih}^* c_{ig,j_ih}
O^\mathrm{s,E}_{ig',ig} .
\end{eqnarray}
Here the norm is $Z' \equiv
\left< \psi_\mathrm{tot} | \psi_\mathrm{tot} \right>$,
$j_i$ is defined such that
$ E_{j_i}^\mathrm{r} = E_\mathrm{tot} - E_i^\mathrm{s}$,
and the energy conservation condition, Eq.~(\ref{Eq:c=0}), has been used.

One sees how energy conservation and entanglement
have collapsed the original four sums over
independent principle energy states
to a single sum over principle energy states.
Specifically the principle energy states of the sub-system
have at this stage collapsed,
whilst the degenerate energy states of the sub-system
remain in superposition form.

An expression for the reservoir entropy will be required
for the two derivations of the statistical average that follow.
One has the axiomatic result given above that the reservoir
entropy for a reservoir energy macrostate is the logarithm
of the number of degenerate states,
\begin{equation}
S^\mathrm{r}(E^\mathrm{r}_j)
= k_\mathrm{B} \ln N_j^\mathrm{r}
, \mbox{ where }
N_j^\mathrm{r} \equiv  \sum_{h}\!\!^\mathrm{(r)}  .
\end{equation}
This must equal the general thermodynamic result
that, given a total energy $E^\mathrm{tot}$
and a sub-system energy $E^\mathrm{s}$,
the reservoir entropy is\cite{TDSM,NETDSM}
\begin{equation}
S^\mathrm{r}(E^\mathrm{r})
=S^\mathrm{r}(E^\mathrm{tot}-E^\mathrm{s})
= \mbox{const.} - \frac{E^\mathrm{s}}{T} .
\end{equation}
By definition, the sub-system has been assumed to be infinitely smaller
than the reservoir and consequently the Taylor expansion can be terminated
at the first term.

Before giving the formal result for the statistical average
of the expectation value,
an equivalent but perhaps more physical argument is first given.
One can impose the condition on the coefficients
in the entangled expansion that
\begin{equation} \label{Eq:c-cond}
\left|
c_{ig,jh}(E_\mathrm{tot})
\right|
=
\left\{ \begin{array}{ll}
1, & E^\mathrm{s}_i + E^\mathrm{r}_j = E_\mathrm{tot} \\
0, & E^\mathrm{s}_i + E^\mathrm{r}_j \ne E_\mathrm{tot} . \\
\end{array} \right.
\end{equation}
That all the allowed coefficients  have the same magnitude
is a statement of the fact that microstates of the total system
that have the same total energy have equal weight.
Hence all total wave functions $\psi_\mathrm{tot}$
compatible with the energy constraint
must have the same weight.
(See also the more general wave space derivation below.)

This choice means that the non-zero coefficients
are of the form $c_{ig,jh} = e^{ i \alpha_{ig,jh}}$,
with the $\alpha$ being real.
(In essence, this is the so-called EPR state.)
Different total wave functions correspond to
different sets of exponents.
If one were to repeat the expectation value with many
different choice of $\psi_\mathrm{tot}$,
the product of coefficients that appears in Eq.~(\ref{Eq:O(psi-tot)}),
 $c_{ig',j_ih}^* c_{ig,j_ih}$,
would average to zero unless $g'=g$.
This will be explicitly shown in the statistical derivation
in wave space that follows next.

Using successively the two forms for the reservoir entropy above,
as well as the form for the coefficients,  Eq.~(\ref{Eq:c-cond}),
and setting $g'=g$,
the average of the expectation value of the operator,
Eq.~(\ref{Eq:O(psi-tot)}),
becomes
\begin{eqnarray}
\left<  \hat O \right>_\mathrm{stat}
& = &
\frac{1}{Z'}  \sum_{ig}\!\!^\mathrm{(s)}
 e^{ S_{j_i}^\mathrm{r} / k_\mathrm{B}}
O^\mathrm{s,E}_{ig,ig}
\nonumber \\ & = &
\frac{1}{Z}  \sum_{ig}\!\!^\mathrm{(s)}
 e^{ -E_i / k_\mathrm{B}T}
O^\mathrm{s,E}_{ig,ig}
\nonumber \\ & = &
\frac{1}{Z}  \sum_{ig,i'g'}\!\!\!^\mathrm{(s)}
\left\{  e^{ -\hat{\cal H} / k_\mathrm{B}T}  \right\}^\mathrm{s,E}_{i'g',ig}
O^\mathrm{s,E}_{ig,i'g'}
\nonumber \\ & = &
\mbox{Tr} \left\{ \hat \wp \, \hat O \right\} ,
\end{eqnarray}
with $\hat \wp = Z^{-1} e^{ \hat S/k_\mathrm{B} }
= Z^{-1} e^{ - \hat {\cal H}/k_\mathrm{B}T }$
and $Z = \mbox{Tr } e^{ - \hat {\cal H}/k_\mathrm{B}T}$.
The penultimate equality follows because
the energy operator is diagonal in the energy representation,
$\langle \zeta^\mathrm{s,E}_{i'g'} |
\hat{\cal H}
| \zeta^\mathrm{s,E}_{ig} \rangle
= E^\mathrm{s}_i \delta_{i,i'} \delta_{g,g'}$.

The Maxwell-Boltzmann probability operator
arises directly from the sum over the degenerate energy microstates
of the reservoir, which gives the exponential of the reservoir
entropy for each particular sub-system energy macrostate.
This converts directly to the matrix representation
of the Maxwell-Boltzmann probability operator
in the energy representation.
Although  the energy representation was used to derive this result,
the final expression as a trace of the product of the two operators
is invariant with respect to the representation.
\cite{Neumann27,Messiah61,Merzbacher70,Bogulbov82}

A more general statistical derivation
integrates over all possible values of the coefficients
that respect energy conservation.
Again, a uniform weight density is invoked
that reflects that fact that all total wave functions
with the same total energy have equal weight,
$\mathrm{d} \psi_\mathrm{tot} \equiv
\mathrm{d} \underline{\underline c}
\equiv \prod_{kg,jh}
\mathrm{d} c_{kg,jh}^\mathrm{r}  \, \mathrm{d}  c_{kg,jh}^\mathrm{i} $,
with the real and imaginary parts of the coefficient
each belonging to the real line, $\in(-\infty,\infty)$.
Using the fundamental form for the expectation value,
Eq.~(\ref{Eq:O(psi-tot)}),
one obtains for the statistical average
\begin{eqnarray} \label{Eq:O-stat-vN}
\left< \hat O \right>_\mathrm{stat}
& = &
\frac{1}{Z''} \int \mathrm{d} \underline{\underline c} \;
O(\psi_\mathrm{tot})
\nonumber \\ & = &
\frac{1}{Z''}
\sum_{i,g,g'}\!\!\!^\mathrm{(s)} \sum_{h=1}^{N_{j_i}^\mathrm{r}}
O^\mathrm{s,E}_{ig',ig}
\int \mathrm{d} \underline{\underline c} \;
c_{ig',j_ih}^* \, c_{ig,j_ih}
\nonumber \\ & = &
\frac{1}{Z''} \sum_{i,g}\!\!^\mathrm{(s)} \sum_{h=1}^{N_{j_i}^\mathrm{r}}
O^\mathrm{s,E}_{ig,ig}
\int \mathrm{d} \underline{\underline c} \; | c_{ig,j_ih}|^2
\nonumber \\ & = &
\frac{\mbox{const.}}{Z''}
\sum_{i,g}\!\!^\mathrm{(s)}
e^{ S_{j_i}^\mathrm{r} / k_\mathrm{B}}
O^\mathrm{s,E}_{ig,ig}
\nonumber \\ & = &
\frac{1}{Z}  \sum_{ig}\!\!^\mathrm{(s)}
 e^{ -E_i^\mathrm{s} / k_\mathrm{B}T}
O^\mathrm{s,E}_{ig,ig}
\nonumber \\ & = &
\mbox{Tr} \left\{ \hat \wp \, \hat O \right\} .
\end{eqnarray}
The third equality follows because the terms
in which the integrand  is odd vanish.
The fourth equality follows because all the integrations
give the same value irrespective of the indeces.
That this value is infinite is of no concern because
it is incorporated into the normalization factor.

The origin here of the Maxwell-Boltzmann probability operator
is the same as in the preceding approximate derivation:
it comes first from the reservoir entropy due to the
sum over degenerate reservoir energy microstates,
and then from exploiting the diagonal nature
of the energy operator in the energy representation.

Although this expression is identical
with the von Neumamn expression,
no reference to an ensemble of systems is made or implied here.
This is also the case in the author's approach
to classical equilibrium\cite{TDSM}
and non-equilibrium\cite{NETDSM} statistical mechanics.

One sees in this derivation that it is the statistical average
that causes the superposition of the degenerate energy microstates
of the sub-system to collapse into a mixture.
(The collapse of the energy macrostates occurred
at the level of the expectation value
due to the energy conservation law and the
entanglement of the reservoir and the sub-system.)
Whether one says that this represents the wave function collapse,
or whether one says that it shows only that the superposition states
do not contribute to the statistical average,
is a moot point.
In the sense that all macroscopic observations and measurements
involve thermodynamic systems
and statistical averages,
either interpretation is compatible with the classical result
that a superposition of states is not observed or measured.

In order to see clearly that the von Neumann expression
for the statistical average, Eq.~(\ref{Eq:O-stat-vN}),
really is akin to a classical average over one state at a time
rather than
a quantum average over superposition states,
one only has to contrast it with the latter,
which would be an integral over
the sub-system wave space
of a probability density $\wp(\psi^\mathrm{s})$
and the expectation value $O(\psi^\mathrm{s})$,
\begin{equation}
\left< \hat O \right>_\mathrm{stat}
\ne
\int \mathrm{d} \psi^\mathrm{s} \; \wp(\psi^\mathrm{s})
O(\psi^\mathrm{s}) .
\end{equation}
This average evidently includes all superposition states of the sub-system.
One can show that it is not possible to reduce it to
the von Neumann expression, Eq.~(\ref{Eq:O-stat-vN}),
without explicitly invoking the collapse of the wave function
into pure energy quantum states.
In contrast, one can show explicitly
(see Paper II in this series) that the average can be written
\begin{equation}
\left< \hat O \right>_\mathrm{stat}
=
\int \mathrm{d} \psi^\mathrm{s} \;
\frac{\langle \psi^\mathrm{s}|\hat \wp \, \hat O | \psi^\mathrm{s}\rangle
}{\langle \psi^\mathrm{s} | \psi^\mathrm{s}\rangle} ,
\end{equation}
which is really just the continuum analogue of the trace,
and as such it manifests the collapse of the wave function.

                \section{Conclusion}

The specific motivation for this paper has been
to derive from first principle an expression for the probability operator
and an expression for the statistical average
of an observable operator in a quantum mechanical system.
Invoking the axiom that all microstates
(i.e.\ states with no further degeneracy)
have the same weight,
it was shown that for a canonical equilibrium system
(i.e.\ where a sub-system can exchange energy with a reservoir),
the probability operator was the Maxwell-Boltzmann operator.
In detail,
the Maxwell-Boltzmann operator ultimately arose
from the sum over the degenerate reservoir energy states
for a reservoir energy implied by the given sub-system energy
(i.e.\ compatible reservoir microstates).
Of course the Maxwell-Boltzmann operator has always been
used as the von Neumann density operator for a canonical system.
The difference here is that the result was derived
from first principles,
not simply invoked by analogy with the classical result.

It was also shown that the statistical average
reduced to the von Neumann form of
the trace of the product of the probability operator
and the observable operator.
This form has the interpretation that the system
has collapsed into a mixture of quantum states
from the underlying superposition of quantum states.
In the present paper the collapse occurred in two stages:
first the principle energy states collapsed
 due to the conservation of energy
and the consequence entanglement of the wave functions
of the sub-system and the reservoir.
And second the contribution from the superposition
of the sub-system energy microstates
(i.e.\ the distinct states of a given energy)
vanished upon statistical averaging the total entangled wave function
over the total Hilbert space.

That the statistical average of an operator on an open system
(i.e.\ a sub-system that can exchange a conserved quantity with a reservoir)
collapses from the expectation value of a superposition of states
to an average over a mixture of states
occurs in any statistical system, not just the present canonical one,
and it has broad implication  beyond quantum statistical mechanics itself.
Specifically there is the measurement problem,
wherein the application of an operator on a system
in a superposed state returns a single eigenvalue of the operator
and collapses the system into the corresponding eigenstate.
And there is the problem of the classical world,
where macroscopic objects are observed to be in only one state at a time
and never in a superposition of states.

It is these two issues
(rather than quantum statistical mechanics \emph{per se})
that have provided the motivation for pursuing the theory
of decoherence by einselection.
The interpretation of them in terms of einselection
have been discussed in detail by the advocates of the latter.
\cite{Zurek82,Joos85,Zurek91,Zurek94,Zeh01,Zurek03,Zurek04,Schlosshauer05}
An alternative interpretation of them based on the present results
is now given.

\subsubsection*{Detecting Schr\"odingers Cat}

The relevance of the present derivation of quantum statistical mechanics
to the measurement problem
is that in general a measurement device has to interact
with a sub-system, generally via the  exchange of energy
or of some other conserved quantity.
As such the measurement device may be regarded as a reservoir
for the sub-system that becomes entangled with it
via the conservation law. The present formalism shows
how this causes the wave function of the sub-system to collapse.


\subsubsection*{Dissecting Schr\"odingers Cat}

The classical nature of the macroscopic world
is elucidated by addressing several questions
that are provoked by the present analysis.
Is it necessary actually to apply an external operator
in order to collapse the wave function
and thereby to give a classical macroscopic system?
The alternative possibility is that
the system spontaneously collapses (e.g.\ due to self-interactions).
In either case,
having collapsed,
does the system remain in the collapsed state between measurements
(i.e.\ is the collapse real and irreversible)?
In general quantum systems are not real in the sense that
their properties have no existence independent of measurement.
Classical systems are the opposite.

1. A system in a mixed state has collapsed
and the statistical average is the same as the classical one
(since the trace becomes an integral over phase space,
the probability operator becomes the probability density,
and the observable operator becomes the corresponding
classical function of phase space).
One is therefore justified in saying
that a collapsed system is a classical system.

2. It was shown in the text that any measurement
on a sub-system able to exchange a conserved variable with a reservoir
revealed a collapsed sub-system.
By the above, this is a classical system.
Since this holds for any measurement (i.e.\ any applied operator)
one is justified in saying that it was not the measurement
that caused the sub-system to collapse but rather
that it was the ability to exchange a conserved variable
that caused the collapse.
There are two reasons for this conclusion.
First, the collapse and subsequent classical average occurs
for any applied operator. If it depended upon the application of an operator,
then it would be sensitive to the particular operator that was applied.
Second,
there is no measurement that could ever be made on the sub-system
that would yield other than the classical result.
The simplest picture, then, is
that the sub-system's collapse was self-induced
(i.e.\ was due to the ability to exchange conserved quantities),
and that this collapse is a real property of the sub-system
that persists between measurements.
Since no  experiment or measurement
could  ever contradict the results of this picture,
by Occam's razor it is the only acceptable picture.
One must conclude that
for a sub-system exchanging with a reservoir,
the collapse is self-induced, irreversible, and real.

3. To address in general the classical nature of a macroscopic object,
one can mentally dissect it into small parts.
Each individual part is a sub-system
that is able to exchange conserved variables
such as energy with the remainder.
Therefore the remainder can
be considered a reservoir for each part.
By the conclusion drawn in the preceding paragraph,
each individual part is therefore in a collapsed state
and is therefore classical.
Since each part of the object is classical,
the object as a whole must be classical.

4. It follows from this that quantum systems are rather special:
they must be prevented from exchanging conserved variables
both with their surroundings and also within themselves.
They must have this property of isolation
or weak linkage in order to maintain their coherent nature.
It is not sufficient to isolate a system from its surroundings
for it to remain quantum;
the components of the system must also be isolated from each other.

Of course just how big an internally interacting, externally isolated
system can get before it collapses and the time scales for the collapse
are quantitative questions that depend upon the details of the system
and are beyond the scope of the present article.
In any case one interesting conclusion from this
is that the wave function of the universe as a whole
must be in a collapsed state.
Quantum superposition states are only possible
for sub-microscopic isolated systems
with weak or non-existent internal interactions.

\textbf{Disclaimer.}
No cats were harmed prior
to measurements performed for this article.


\appendix
                \section{Environmental Selection} \label{Sec:ein}

Environmental selection or einselection
\cite{Zurek82,Joos85,Zurek91,Zurek94,Zeh01,Zurek03,Zurek04,Schlosshauer05}
refers
to the purification of a sub-system from a superposed state
to a mixture of pure quantum states
by its interaction with the reservoir.
It  is based  on the notion that
in a Hilbert space of large dimension, say with $M$ degrees of freedom,
two normalized wave functions chosen at random
are almost certainly orthogonal.
One expects that on average their relative scalar product would go like
\begin{equation}
 \left< \psi | \phi \right>
\sim c M^{-1/2}
\approx \delta(\psi-\phi).
\end{equation}
The stochastic variable $c$ is order unity
and is on average zero.
Of course if $\psi=\phi$, this would equal unity.

To see how einselection works in detail,
suppose that in some orthonormal basis
one has superposed wave functions of the isolated sub-system,
\begin{equation}
\left| \psi  \right> = \sum_n  a_n   \left| \zeta_n  \right>
, \mbox{ and }
\left| \phi  \right> = \sum_n  b_n   \left| \zeta_n  \right> .
\end{equation}
The scalar product of these is
\begin{equation}
\left< \psi | \phi \right>
=
\sum_{m,n} a_m^*  b_n \left< \zeta_m | \zeta_n \right>
= \sum_{n} a_n^*  b_n ,
\end{equation}
and the expectation value of some observable is
\begin{equation}
O(\psi) \equiv
\left< \psi \right| \hat O \left| \psi \right>
=
\sum_{m,n} a_m^*  a_n  O_{mn} ,
\end{equation}
assuming $\psi$ is normalized,
$\sum_n a_n^*  a_n =1$.

If the sub-system is allowed to interact with the reservoir,
then the total system Hilbert space is the direct product of the
Hilbert spaces of the sub-system and the reservoir,
and it is conventionally assumed that the total wave function may be written
$  \left| \psi \right>_\mathrm{s} \left| \varepsilon \right>_\mathrm{r}
\equiv \left| \psi,\varepsilon \right>  $.
\cite{Zurek82,Joos85,Zurek91,Zurek94,Zeh01,Zurek03,Zurek04,Schlosshauer05}
Note that in contrast to the analysis in the text
that accounted for the conservation laws,
this is a product state, not an entangled state.

Suppose that  the reservoir is initially in the normalized state
$ \left| \varepsilon_0 \right>_\mathrm{r}$.
After time $t'$,
suppose  that the sub-system wave function evolves to
$\left| \psi \right>_\mathrm{s}
\stackrel{t'}{\Rightarrow} \left| \psi' \right>_\mathrm{s} =
\sum_n a_n'  \left| \zeta_n \right>_\mathrm{s}$.
Also suppose that when the sub-system is in the state
$\left| \zeta_n \right>_\mathrm{s}$,
the initial reservoir wave function evolves to
$ \left| \varepsilon_0 \right>_\mathrm{r}
\stackrel{t'}{\Rightarrow}  \left| \varepsilon_n' \right>_\mathrm{r}$.
The subscript $n$ occurs because this evolution
is affected by the state of the sub-system.
By the quantum linear superposition principle,
the total scalar product evolves to
\begin{eqnarray}
\left< \varepsilon_0 , \psi  |  \phi ,\varepsilon_0 \right>
& = &
\sum_{m,n}  a_m^*   b_n
\left< \varepsilon_0 , \zeta_m   |   \zeta_n , \varepsilon_0 \right>
\nonumber \\ & \stackrel{t'}{\Rightarrow} &
\sum_{m,n}  a_m'^{*}   b_n'
\left< \varepsilon_m' , \zeta_{m}   |   \zeta_{n} , \varepsilon_n' \right>
\nonumber \\ & = &
\sum_{m,n}  a_m'^{*}   b_n'  c_m' \delta_{m,n} \delta_{m,n}
\nonumber \\ & = &
\sum_{m}  a_m'^{*}  b_m'   c_m'
\nonumber \\ & = &
\sum_{m}  a_m^{*}  b_m     .
\end{eqnarray}
Here $ \left< \varepsilon_m'    |  \varepsilon_n' \right>_\mathrm{r}
= c_m' \delta_{m,n}$,
because  due to their interaction,
the individual evolution operators on each space are not unitary;
the $\varepsilon_n' $ may be approximately orthogonal
but they do not necessarily remain normalized,
as is reflected in the constant $c_m'$.
However because the time propagator for the total system is unitary,
it preservers scalar products
and so one must have
$a_m'^{*}   b_m'  c_m'
= a_m^* b_m \left< \varepsilon_0    |  \varepsilon_0 \right>_\mathrm{r}
= a_m^* b_m $.
This holds term by term because of the linear superposition principle.
One sees that the scalar product of wave functions
of the sub-system interacting with the reservoir
are unchanged  from that of the isolated system.

In contrast,
the expectation value of an arbitrary observable
acting only on the sub-system is
\begin{eqnarray} \label{Eq:ein-tr}
\left< \varepsilon_0 , \psi \right| \hat O \left| \psi, \varepsilon_0 \right>
& = &
\sum_{m,n}  a_m^*   a_n
\left< \varepsilon_0 , \zeta_m  \right| \hat O \left|
\zeta_n , \varepsilon_0 \right>
\nonumber \\ & \stackrel{t'}{\Rightarrow} &
\sum_{m,n}  a_m'^*   a_n'
\left< \varepsilon_m' , \zeta_{m}
\right| \hat O \left|  \zeta_{n} , \varepsilon_n' \right>
\nonumber \\ & = &
\sum_{m,n}  a_m'^*  a_n'  O_{mn}  c_m' \delta_{m,n}
\nonumber \\ & = &
\sum_{m}  a_m^*   a_m O_{mm}   .
\end{eqnarray}
This is different to what was obtained for a sub-system alone.
It is the expectation value for a mixture of pure quantum states
rather than that of a superposition of states.
This is the result and meaning of einselection:
the interaction with the reservoir
results in a loss of coherence of the original wave function
and transforms it from a superposition of states
into a weighted mixture of pure quantum states.

\subsubsection*{Discussion}

The approach taken  in the present paper
is obviously rather similar to this einselection mechanism for decoherence,
\cite{Zurek82,Joos85,Zurek91,Zurek94,Zeh01,Zurek03,Zurek04,Schlosshauer05}
in that the collapse of the system ultimately arises from the interactions
between the sub-system and the reservoir.

The main difference between the two approaches
is that  einselection decoherence
takes the sub-system and its environment to form a product state,
whereas in the text it is shown that they must form an entangled state,
and it is this entanglement and conservation law
that specifically leads to the decoherence.
A second difference is that the einselection expectation value,
Eq.~(\ref{Eq:ein-tr}) is a sum over a single index
of only the diagonal terms of both
the operator matrix and the dyadic density matrix
in an arbitrary basis,
whereas the von Neumann trace, Eq.~(\ref{Eq:O-stat-vN}),
is in general a sum over all components of the two matrices
that only reduces to a single sum of diagonal terms
in the operator basis
or in the entropy (here energy) basis.


Of course one very nice feature of the einselection approach to decoherence
is that it provides a detailed physical picture
of the evolution of the collapse of the wave function.
Despite the differences in detail,
the conclusions from the present approach
and from the einselection approach
are the same:
wave function collapse ultimately arises from the interactions
between the sub-system and the environment or reservoir.



                \section{Uniform Weight Density} \label{Sec:weight}

Schr\"odinger's equation for an isolated system is
\begin{equation}
i \hbar \left| \right.\! \dot \psi \!\left. \right>
=
\hat{\cal H} \left| \psi \right>
, \mbox{ or }
i \hbar  \underline{\dot \psi}
=
\underline{\underline{\cal H}} \cdot \underline{\psi} ,
\end{equation}
in some arbitrary representation for the matrix form.
One can construct a trajectory from this,
$\psi(t|\psi_0,t_0)$.

Associated with each point in wave space is a weight density $\omega(\psi,t)$,
which when normalized is the probability density $\wp(\psi,t)$.
The quantity $\mathrm{d} \psi \, \wp(\psi,t)$
gives the probability of the system being within $|\mathrm{d} \psi|$
of $\psi$.
As a probability the product is a real non-negative number:
$\mathrm{d} \psi \, \wp( \psi,t)
=
\left| \mathrm{d} \psi \, \wp( \psi,t) \right|
=
\left| \mathrm{d} \psi  \right| \,
\left|  \wp( \psi,t) \right|$.
Since it is always the product that occurs,
without loss of generality one may take each to be individually real.
The normalization is
\begin{equation}
\int \mathrm{d} \psi \, \wp( \psi,t) \,
= \int \mathrm{d} \underline \psi \, \wp(\underline \psi,t) \,
= 1 .
\end{equation}In the matrix representation,
$\underline \psi = \{ \psi_n \}$, $n=1,2,\ldots $.
Since the coefficients are complex,
 $\psi_n = \psi_n^\mathrm{r} + i \psi_n^\mathrm{i} $,
one can write the infinitesimal volume element as
\begin{equation} \label{Eq:dpsi}
\mathrm{d} \underline \psi
=
\mathrm{d} \underline \psi^\mathrm{r}\,
\mathrm{d}\underline \psi^\mathrm{i} \,
\equiv
\mathrm{d}\psi_1^\mathrm{r}\, \mathrm{d}\psi_1^\mathrm{i} \,
\mathrm{d}\psi_2^\mathrm{r}\, \mathrm{d}\psi_2^\mathrm{i}
\ldots
\end{equation}
With this the integrations are over the real line,
$\psi_n^\mathrm{r} \in [-\infty,\infty]$, and
$\psi_n^\mathrm{i} \in [-\infty,\infty]$.

The compressibility of the trajectory is
\begin{eqnarray}
\frac{\mathrm{d} \dot \psi}{\mathrm{d} \psi }& = &
\underline \partial_{\psi^\mathrm{r}}
\cdot  \dot{ \underline \psi}\,\!^\mathrm{r}
+ \underline \partial_{\psi^\mathrm{i}}
\cdot  \dot{ \underline \psi}\,\!^\mathrm{i}
\nonumber \\ & = &
\underline \partial_\psi \cdot  \dot{ \underline \psi}
+ \underline \partial_{\psi^*} \cdot  \dot{ \underline \psi}\,\!^*
\nonumber \\ & = &
\frac{1}{i\hbar} \mbox{Tr }\underline{ \underline { \cal H}}
-
\frac{1}{i\hbar} \mbox{Tr }\underline{ \underline { \cal H}}
\nonumber \\ & = & 0 .
\end{eqnarray}
(The first two equalities are general;
the final two hold for Schr\"odinger's equation.)

The compressibility
gives the logarithmic rate of change of the volume element.
\cite{TDSM,NETDSM}
Hence from the vanishing of the compressibility for the
adiabatic Schr\"odinger equation
one sees that the volume element
is a constant of the motion of the isolated system,
\begin{equation} \label{Eq:|dpsi|}
 \mathrm{d} \psi(t) = \mathrm{d} \psi_0 .
\end{equation}

The total time derivative of the probability density is
\begin{eqnarray} \label{Eq:dpsi/dt}
\frac{\mathrm{d}  \wp(\psi,t) }{\mathrm{d} t}
& = &
\frac{\partial \wp(\psi,t) }{\partial t}
+
\dot {\underline \psi}\,\!^\mathrm{r} \cdot
\underline \partial_{ \psi^\mathrm{r}} \wp(\psi,t)
+
\dot {\underline \psi}\,\!^\mathrm{i} \cdot
\underline \partial_{ \psi^\mathrm{i}} \wp(\psi,t)
\nonumber \\ & = &
\frac{\partial \wp(\psi,t) }{\partial t}
+
\dot {\underline \psi} \cdot
\underline \partial_{ \psi} \wp(\psi,t)
+
\dot {\underline \psi}\,\!^* \cdot
\underline \partial_{ \psi^*} \wp(\psi,t)
\nonumber \\ & = &
\frac{\partial \wp(\psi,t) }{\partial t}
+
\underline \partial_{\psi} \cdot
\left[ \dot {\underline \psi}  \wp(\psi,t) \right]
\nonumber \\ && \mbox{ }
+
\underline \partial_{ \psi^*}
\cdot \left[ \dot {\underline \psi}\,\!^*  \wp(\psi,t) \right] .
\end{eqnarray}
The first and second equalities are definitions that hold in general.
The final equality, which uses the vanishing of the compressibility,
is valid for an isolated system
that evolves via Schr\"odinger's equation.
One can identify from this
$J_\wp(\psi,t) \equiv \dot { \psi}  \wp(\psi,t)$
as the probability flux.

In view of the deterministic nature of  Schr\"odinger's equation,
the evolution of the probability density is given by
\begin{equation}
\wp( \psi_1,t_1)
=
\int \mathrm{d} \psi_0 \,
\wp( \psi_0,t_0)\,
\delta \! \left( \psi_1
-  \psi(t_1| \psi_0,t_0) \right) .
\end{equation}
Using the constancy of the volume element,
$\mathrm{d}  \psi_0 = \mathrm{d}  \psi_1 $,
setting $t_1=t_0+\Delta_t$,
and expanding to linear order in $\Delta_t \rightarrow 0$,
one has
\begin{eqnarray}
\lefteqn{
\wp( \psi_1,t_0)
+ \Delta_t \frac{\partial \wp( \psi_1,t_0) }{\partial t_0}
}   \\ \nonumber
& = &
\int \mathrm{d} \psi_1 \,
\wp( \psi_0,t_0)\,
\delta\!\left(  \psi_1  - \psi(t_1| \psi_0,t_0)
\right)
 \nonumber \\ & = &
\wp\!\left( \psi_1
-  \Delta_t \dot{  \psi} ,t_0\right)
 \nonumber \\  & = &
\wp( \psi_1,t_0)
- \Delta_t  \dot{ \underline \psi} \cdot
\underline \partial_{\psi} \wp( \psi,t_0)
- \Delta_t  \dot{ \underline \psi}\,\!^* \cdot
\underline \partial_{\psi^*} \wp(\psi,t_0) .
 \nonumber 
\end{eqnarray}
Hence the partial time derivative is
\begin{equation} \label{Liou-isol}
\frac{\partial \wp(\psi,t) }{\partial t}
=
-\dot {\underline \psi} \cdot \underline \partial_{\psi^*} \wp(\psi,t)
- \dot {\underline \psi}\,\!^* \cdot \underline \partial_{\psi^*} \wp(\psi,t) .
\end{equation}
This result is valid for an isolated system
that evolves under Schr\"odinger's equation.

Inserting this into the second equality in Eq.~(\ref{Eq:dpsi/dt}),
it may be seen that the total time derivative
of the probability density  vanishes,
$\mathrm{d} \wp(\psi,t)/\mathrm{d} t = 0$.
Hence for an isolated system,
the probability density is a constant of the motion,
\begin{equation} \label{Eq:psi(t)}
\wp(\psi(t),t) = \wp(\psi_0,t_0) ,
\end{equation}
where $\psi(t) \equiv \psi(t|\psi_0,t_0)$.

In an equilibrium system,
(i.e.\ the Hamiltonian operator does not depend upon time),
the probability density
is not explicitly dependent on time,
so that one can write $\wp( \psi )$.
In this case the result becomes
\begin{equation}
\wp( \psi(t))
=
\wp( \psi_0) .
\end{equation}

Finally, one may invoke what might be called
the weak form of the classical ergodic hypothesis,
which says  that a single trajectory passes sufficiently close
to all relevant points of the state space.
This means that any two  points in state space,
$ \psi_1$ and $ \psi_2$,
with the same norm and energy but otherwise arbitrary,
lie on a single trajectory.
By the above, they therefore have the same probability density
\begin{equation}
\wp( \psi_2) = \wp( \psi_1),
\end{equation}
if $ E( \psi_2) =  E( \psi_1) $ and $N( \psi_2) =  N( \psi_1)$.

One possible objection
to the weak form of the ergodic hypothesis
(that the trajectory visits all points on the energy hypersurface)
is that an isolated system always remains in its initial
energy microstate or  superposition of energy microstates,
since these are eigenstates of the energy operator.
However, if the wave function is represented in some other basis,
then the concept of a trajectory through wave space $\psi(t)$ is meaningful,
(successive measurements with the same operator do not necessarily
yield the same value)
and the  weak form of the ergodic hypothesis is plausible
and leads to the conclusion that the weight density
of wave space is uniform
(on a constant energy and magnitude hypersurface).

Using this result the weight density of state space
must be of the form
$\omega(\psi) = \omega(E(\psi))$.
(For simplicity the magnitude is neglected since it
only trivially affects the analysis.)
The only time one needs to compare
the weight densities of points with different energies
is when energy exchange with a reservoir occurs.
In such a case the total weight density must be of the form
$\omega(\psi_\mathrm{tot})
= \omega(E_\mathrm{s}(\psi_\mathrm{tot}))
\omega(E_\mathrm{tot}-E_\mathrm{s}(\psi_\mathrm{tot}))$,
which has solution
$\omega(\psi) = w e^{\alpha E(\psi)}$,
with $w$ and $\alpha$ arbitrary constants.
This means that any variation in the weight density of wave space
with energy of the sub-system will cancel identically
with the equal and opposite dependence
of the weight density of the reservoir
on the sub-system energy.
(This is a true statement for each individual point
in the wave space of the reservoir.
Summing over the reservoir degeneracy remains to be carried out,
as in the text.)
Hence without loss of generality one can set $\alpha = 0$
and the weight density may be set to unity,
\begin{equation}
\omega( \psi) = 1 .
\end{equation}
This result applies to the whole of state space of an isolated system,
not just to the hypersurface to which Schr\"odinger's trajectory
is constrained in any given case.
By design, this is real and positive.
The logarithm of this gives the internal entropy of the wave states,
\begin{equation} \label{Eq:Ss=0}
S( \psi) = 0 .
\end{equation}

\subsubsection*{Alternative Derivation}

An arguably more satisfactory derivation is as follows.
The magnitude of the wave function is redundant in the sense that
expectation values only depend upon the normalized wave function.
For example, denoting the latter as
$\tilde \psi \equiv N(\psi)^{-1/2} \psi$,
with the magnitude being $ N(\psi) \equiv \left< \psi | \psi \right>$,
the energy is
$E(\psi) =  \left<\!\right. \tilde \psi
| \hat {\cal H} | \tilde \psi \left.\! \right>$.
In addition, since the isolated system is confined to a hypersurface
of constant magnitude $N$ and energy $E$,
one can denote a representation of that particular Hilbert sub-space
by $\chi$.
In this notation the normalized wave function is
$\tilde \psi = \tilde\psi(E, \chi)$,
and the full wave function is
$\psi =  \psi(N,\tilde\psi) = \psi(N, E, \chi)$.

Now the weight density on the hypersurface,
$\omega(\chi|NE)$ will be derived,
and then this will be transformed to the weight density
on the full wave space.
The derivation is the quantum analogue
of that given for classical statistical mechanics
in \S 5.1.3 of Ref.~\onlinecite{TDSM}.
The axiomatic starting point is that the fundamental statistical average
is a simple time average.
The implication of this is that the weight is uniform in time,
which is to say that it must be inversely proportional to the speed,
\begin{eqnarray}
\omega(\chi|NE) & \propto & | \dot \psi |^{-1}
\nonumber \\ & = &
\left<\!\right.  \psi | \hat {\cal H} \hat {\cal H}
|  \psi \left.\! \right>^{-1/2} .
\end{eqnarray}
This is just the time that the system spends in a volume element
$|\mathrm{d} \chi|$.
The proportionality factor is an immaterial constant.
This weight density is now successively transformed to the full wave space.
For this one requires in turn the  Jacobean
for the transformation $\chi \Rightarrow \tilde \psi$,
\begin{equation}
\left| \tilde \nabla E \right|
=
\left[ \frac{\partial E(\psi)
}{\partial \left| \!\right.  \tilde \psi \left.\! \right> }
\frac{\partial E(\psi)
}{\partial \left< \!\right.  \tilde \psi \left.\! \right| }
\right]^{1/2}
=
\left<\!\right.  \tilde \psi | \hat {\cal H} \hat {\cal H}
|  \tilde \psi \left.\! \right>^{1/2} ,
\end{equation}
and for the transformation $\tilde \psi \Rightarrow \psi$,
\begin{equation}
\left|  \nabla N \right|
=
\left[ \frac{\partial N(\psi)
}{\partial \left| \!\right.   \psi \left.\! \right> }
\frac{\partial N(\psi)
}{\partial \left< \!\right.   \psi \left.\! \right| }
\right]^{1/2}
=
\left<   \psi | \psi \right>^{1/2} .
\end{equation}
With these  the full weight density is
\begin{eqnarray}
\omega(\psi)
& = &
\omega(\chi|NE) \,
| \tilde \nabla E | \,
|  \nabla N |
\nonumber \\ & \propto &
\frac{\left<\!\right.  \tilde \psi | \hat {\cal H} \hat {\cal H}
|  \tilde \psi \left.\! \right>^{1/2}
\left<   \psi | \psi \right>^{1/2}
}{\left<\!\right.  \psi | \hat {\cal H} \hat {\cal H}
|  \psi \left.\! \right>^{1/2} }
\nonumber \\ & = &
1 .
\end{eqnarray}
The interpretation of this is that the weight density
is  inversely proportional to the speed of the trajectory
(i.e.\ large speed means less time per unit volume),
and linearly proportional to the number of hypersurfaces
that pass through each volume of wave space
(i.e.\ for fixed spacing between discrete hypersurfaces,
$\Delta_E$ and $\Delta_N$,
large gradients correspond to more hypersurfaces
per unit wave space volume).

Just as in the classical case for Hamilton's equations of motion,\cite{TDSM}
it is a remarkable consequence
of Schr\"odinger's equation that the speed
is identical to the magnitude of the gradient of the hypersurface,
so that these two cancel to give a weight density
that is uniform in wave space.

An objection to the uniform weight density of wave space may be raised,
namely that a non-linear transformation of wave space
would lead to a non-uniform weight in the transformed coordinates.
However, any such non-linear transformation of the wave function
would destroy the linear homogeneous form
for the operator equations of quantum mechanics.
In this sense the form of Schr\"odinger's equation
determines uniquely that the weight of wave space should be uniform.

Both derivations
(the uniformity of the probability density along a trajectory,
and the proportionality of weight to time and gradients)
yield the same result,
namely that the weight density is uniform in wave space.
Since the microstates of an operator are just its eigenfunctions
in wave space,
$\left| \!\right. \zeta_{kg}^O \left. \! \right>$,
this result implies that these microstates all have equal weight,
which is the axiomatic starting point taken in the main text.


\begin{thebibliography}{99}
\bibitem{FN2}
An alternative interpretation of this result is that in those
sub-systems in which the energy macrostates are so spaced
as to preclude energy exchange with a reservoir,
entanglement and the consequent collapse of the wave function
may not occur.
\end{thebibliography}
\end{document}